\documentclass[aps,prl,twocolumn,showpacs,superscriptaddress,floatfix]{revtex4}
\usepackage{graphics}
\usepackage{epsfig}
\usepackage{subfigure}
\usepackage{amsmath,amsfonts,amssymb,graphicx}

\begin{document}

\title{Degree Landscapes in Scale-free Networks}

\author{Jacob Bock Axelsen}\email{bock@nbi.dk}
\affiliation{NBI, Blegdamsvej 17, Dk 2100, Copenhagen, Denmark}
\author{Sebastian Bernhardsson}
\affiliation{Department of Theoretical Physics, Ume{\aa} University,
901 87 Ume{\aa}, Sweden}
\author{Martin Rosvall}
\affiliation{Department of Theoretical Physics, Ume{\aa} University,
901 87 Ume{\aa}, Sweden}
\author{Kim Sneppen}
\affiliation{NBI, Blegdamsvej 17, Dk 2100, Copenhagen, Denmark}
\author{Ala Trusina}
\affiliation{NBI, Blegdamsvej 17, Dk 2100, Copenhagen, Denmark}

\date{\today}

\begin{abstract}
We generalize the degree-organizational view
\cite{Pastor-Satorras,maslov2002b,maslov2002,newman2002,trusina} of
real-world networks with broad degree-distributions
\cite{albert99,albert-review,dorogovtsev-review,newman-review} in a
landscape analogue with mountains (high-degree nodes) and valleys
(low-degree nodes).  For example, correlated degrees between adjacent
nodes corresponds to smooth landscapes (social networks)
\cite{newman2002}, hierarchical networks to one-mountain landscapes
(the Internet) \cite{trusina}, and degree-disassortative networks
without hierarchical features to rough landscapes with several
mountains \cite{Pastor-Satorras,trusina,newman2003a,newman2003b}. We
also generate ridge landscapes to model networks organized under
constraints imposed by the space the networks are embedded in,
associated to spatial or, in molecular networks, to functional
localization. To quantify the topology, we here measure the widths of
the mountains and the separation between different mountains.
\end{abstract}

\maketitle

The broad degree-distribution in many real-world networks makes it
meaningfull to investigate the topological organization of nodes in
terms of their degree.  It has been found that many social networks
are assortative, with correlated degrees of adjacent nodes, but that
technological and biological networks often are disassortative, with
anticorrelated degrees of adjacent nodes
\cite{newman2002,maslov2002,newman2003a,newman2003b}.  The degree
correlation-profile, generated by comparison between the network and
its randomized counterparts without degree correlations, uncovers in
the Internet an overrepresentation of links between intermediate- and
low-degree nodes and a slight overrepresentation of links between the
nodes of highest degrees \cite{maslov2002b}.  Contrary, most
biological networks have an underrepresentation of links between the
hubs \cite{maslov2002}.  To characterize the organization beyond
correlations between adjacent nodes, Trusina \textsl{et
al.~}\cite{trusina} introduced the hierarchy measure ${\cal F}$.
${\cal F} \in ]0,1]$, is the fraction of shortest paths between all
pairs of nodes that are degree hierarchical \cite{Gao}.  The degrees
of the nodes along a degree-hierarchical shortest path are organized
in strictly ascending, strictly descending, or first strictly
ascending and then strictly descending order.  It was found that
biological networks with decentralized hubs stand out from other
networks with a very low value of ${\cal F}$ \cite{trusina}.

\begin{figure}[tp]
\centering
\includegraphics[width=\columnwidth ]{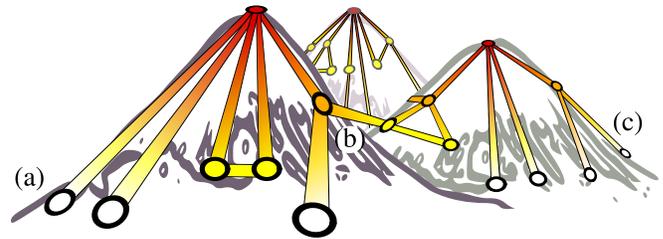}
\caption{A network as a degree landscape with mountains and valleys,
with the altitude of a node proportional to its degree. A route over
one mountain corresponds to making a degree-hierarchical path ((a) to
(b)) while climbing over more than one mountain breaks the
degree-hierarchical path ((a) to (c)).}
\label{fig1}

\end{figure}

Here we generalize the presented findings in a landscape analogue,
with mountains (high degree nodes) and valleys (low degree nodes).
With this interpretation, social networks form smooth landscapes,
the Internet one single mountain with first ascending and then descending hierarchical paths,
whereas biological networks form rough landscapes with several mountains and
broken hierarchical paths.
To quantify the topology and make it possible to compare different networks,
we in this paper measure the typical width of individual mountains
and the separation between different mountains (Fig.\ \ref{fig1}).

\begin{figure*}[!tp]
\centering
\includegraphics[width=\textwidth]{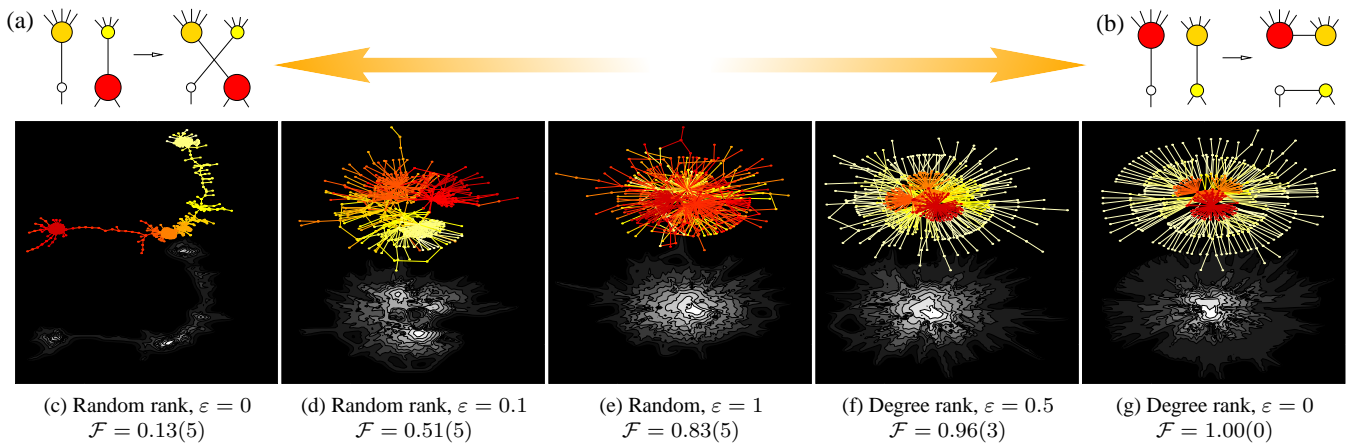}
\caption{Networks as degree landscapes organized from ridge landscapes
(c), via random landscapes (e), to peaked one-mountain landscapes (g).
The links are pairwise swapped to connect high-ranked nodes to
organize the nodes globally according to their rank (color coded from
red for high rank, to white for low rank), with random swaps at
different rates $\varepsilon$.  The rank is set randomly to the nodes,
as in the swap example in (a), in (c-d), and proportional to the
degree of the nodes, as in the swap example in (b), in (f-g).  The
random network in (e) corresponds to $\varepsilon=1$.  The
corresponding degree landscapes are color coded according to altitude
from black (low) to white (high).  The networks are scale-free with an
exponent $\gamma = 2.5$ and of size N=400, originally generated with
the algorithm suggested \cite{trusina}. The layout is generated with
the Kamada-Kawai algorithm in Pajek \cite{pajek}.
\label{fig2}}
\end{figure*}

To complement the methods to generate random networks (random
one-mountain landscapes) \cite{maslov2002,maslov2002b} and completely
hierarchical networks (peaked one-mountain landscapes) \cite{trusina}
with preserved degree-sequences, we here suggest a method to generate
\emph{ridge landscapes} (Fig.\ \ref{fig2}).  In its simplest
implementation, we assign a random rank to every node in a network,
and organize the nodes hierarchically based on their rank.  This
method creates non-random networks, distinguished by a separation of
hubs (disassortative with low ${\cal F}$).  We argue that the ridge
landscapes can represent networks organized under different spatial
constraints put on real-world networks during their evolution.

We start by reviewing the method presented by Trusina \textsl{et
al.~}\cite{trusina} to generate degree-hierarchical networks, here
denoted degree-rank hierarchies.  In the same time, we in detail
present the suggested method to generate ridge landscapes (or
random-rank hierarchies).  The networks evolve by pair-wise rewirings
of the links, with every rewiring constrained by the rank of the nodes
involved in the rewiring.  The rank of a node is proportional to its
degree in the \emph{degree-rank hierarchy}, and set to a random rank
(degree independent) in the \emph{random-rank hierarchy}.  At every
time step, two random links are chosen, and reconnected such that the
two nodes with the highest ranks become adjacent (see Fig.\
\ref{fig2}(a-b)).  In this way the degree of every node is kept
constant and the nodes are globally organized in decreasing
rank-order.  To be able to investigate networks between the
random-rank hierarchical networks respectively the degree-rank
hierarchical networks, and random networks, we allow for random
link-swaps without the constraints set by the rank of the nodes.  A
probability $\varepsilon$ to make a random link-swap corresponds in
this way to an error rate in the creation of the extreme networks.
When $\varepsilon \rightarrow 1$ the methods become equivalent to the
randomization of networks with remained degree-sequence suggested in
\cite{maslov2002}, see Fig.\ \ref{fig2}(e).

Figure \ref{fig2} shows topologies generated with the different
models.  They all originate from a random scale-free network (shown in
Fig.\ \ref{fig2}(e)) with degree distribution, $P(k)\propto k^{-2.5}$
and system size $N=400$, generated with the method suggested in
\cite{trusina}.  The extreme networks, the perfect random-rank
hierarchy in Fig.\ \ref{fig2}(c) and the perfect degree-rank hierarchy
Fig.\ \ref{fig2}(g) ($\varepsilon = 0$), surround the networks with
increasing error-rate towards the random scale-free network with
$\varepsilon = 1$ in the middle (Fig.\ \ref{fig2}(e)).  The perfect
degree-rank hierarchy consists of a tightly connected core of large
degree nodes, that forms a very peaked mountain.  The landscape is not
too far from the, although flatter, random case.  The random-rank
hierarchy, on the other hand, forms a very stringy and non-random
structure --- a ridge landscape.  The length of the string is of the
order $D \propto N$, with very long pathways that break the
small-world property found in most real-world networks. However, as
for the original small-world scenario proposed by \cite{Watts}, the
large diameter of the stringy scale-free networks collapses if small
perturbations exist in the hierarchical organization.  If we generate
the network with a small error rate $\varepsilon$, the diameter of the
network collapses as seen in Fig.~\ref{fig2}(d).  Note that the color
gradient indicates that the random-rank hierarchy is still intact at
this stage, and that the hubs (mountains) are separated.  The
degree-rank hierarchy in Fig.~\ref{fig2}(f) is rewired with a higher
error rate $\varepsilon = 0.5$, while still maintaining a high level
of hierarchical organization.

In both cases, the two organizing principles
leads to higher clustering \cite{Watts}, more triangles,
than in the random counterparts (not shown). This is
expected, as organization along any coordinate tends to make
friends of friends more alike.
The effect is stronger in the degree-rank hierarchy,
since the clustering automatically increases further
when the hubs with their many links are connected.

In Fig.\ \ref{fig2}, we also quantify the degree-hierarchical
organizations of the scale-free networks organized by, respectively,
degree- and number rank. For the random scale-free network with degree
distribution $P(k)\propto k^{-2.5}$ and $N=1000$ nodes, ${\cal
F}=0.83\pm0.05$.  The networks organized hierarchically according to
degree-rank (as in Fig.\ \ref{fig2}(b)) have ${\cal F}=1$ as expected.
Further, when introducing a finite error rate $\varepsilon$ for link
rewirings toward the degree-hierarchy we find that its topology is
robust in the sense that both diameter (not shown) and ${\cal F}$
remain unchanged for even quite large errors.  The perfect random-rank
hierarchy has a much lower degree hierarchical organization, ${\cal
F}=0.13 \pm 0.05$.  Because of the collapsing diameter, the
random-rank hierarchy is not as robust as the degree-rank hierarchy to
errors in the rewiring.

\begin{figure}[tp]
\centering
\includegraphics[width=.9\columnwidth ]{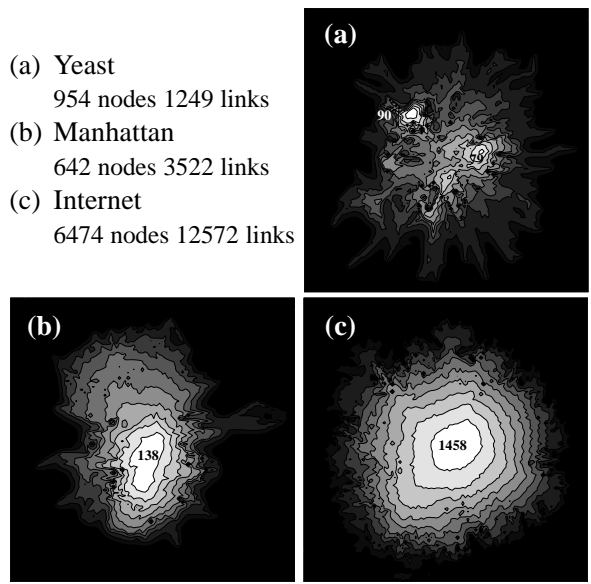}
\caption{ Real-world networks as degree landscapes. The coloring of
the altitudes are relative to the summit altitude.  Yeast in (a) is
the protein-protein interaction network in \emph{Saccharomyces
Cerevisia} \cite{ito}, Manhattan in (b) is the dual map of Manhattan
with streets as nodes and intersections as links \cite{rosvall05}, and
the Internet in (c) is the network of autonomous systems
\cite{internet}.  The topological maps are not based on the real space
the networks are embedded in, but the Kamada-Kawai algorithm in Pajek
\cite{pajek}.
\label{fig3}}
\end{figure}
Figure \ref{fig3} shows, in increasing degree-hierarchical order, a
number of real-world networks as degree landscapes: Yeast in (a) is
the protein-interaction network in \emph{Saccharomyces Cerevisia}
detected by the two-hybrid experiment \cite{ito}, Manhattan in (b) is
the dual map of Manhattan with streets as nodes and intersections as
links \cite{rosvall05}, and the Internet in (c) is the network of
autonomous systems \cite{internet}.  Internet and Manhattan consist of
one single mountain with first ascending and then descending
hierarchical paths, whereas yeast forms a rough landscape with several
mountains and broken hierarchical paths.

\begin{figure}[tp]
\includegraphics[width=\columnwidth]{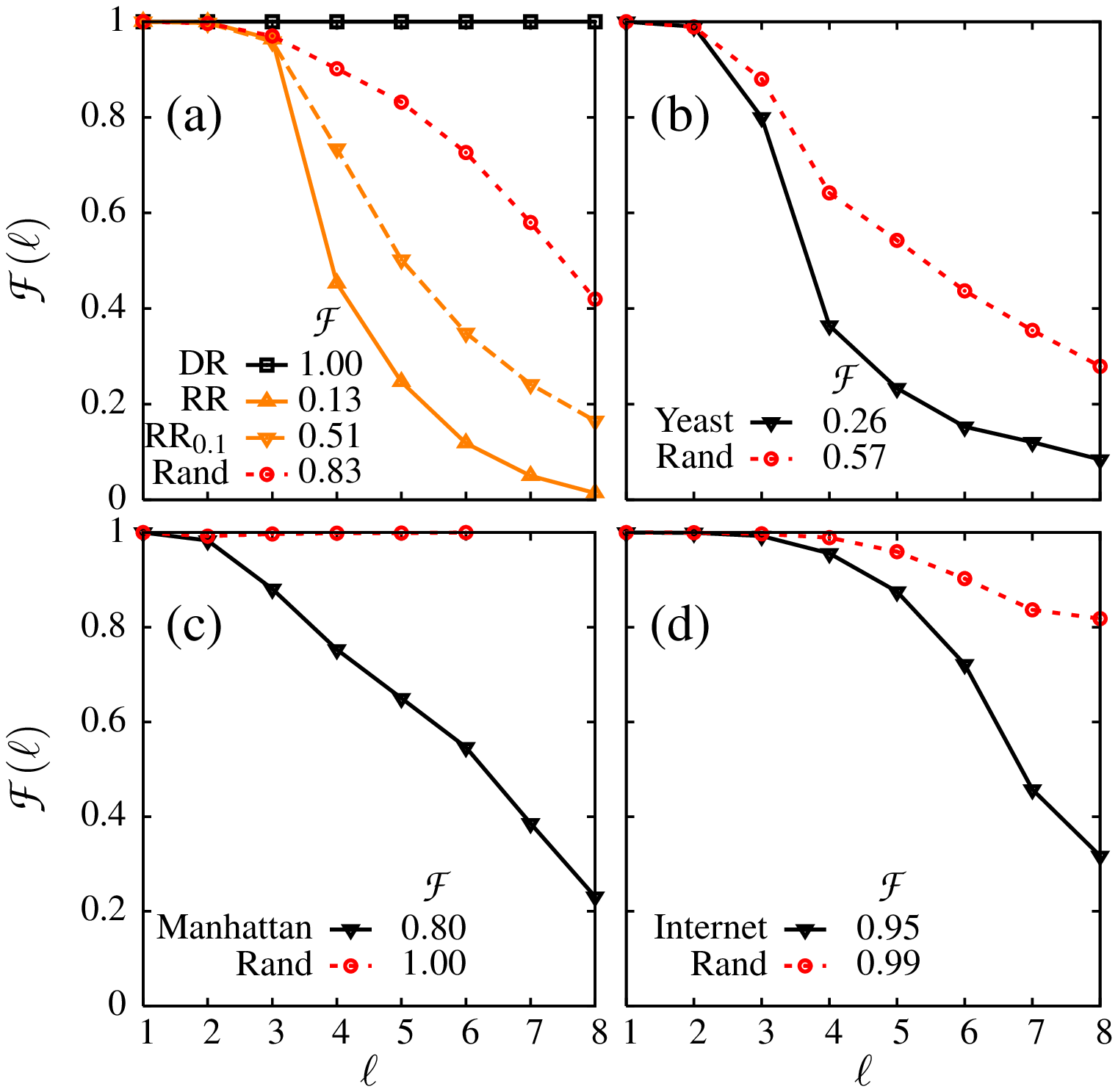}
\caption{ The degree-hierarchical organization as a function of path
length.  ${\cal F}(\ell )$ is the fraction of pair of nodes, separated
by a distance $\ell$, that are connected by a degree-hierarchical
path.  (a) shows the two model networks: The degree-rank hierarchy
(Degree-Rank), the random-rank hierarchy (Random-Rank) for
$\varepsilon=0$ and $0.1$, together with the random scale-free network
(Random).  The real-world networks in (b-d) are the same as in Fig.\
\ref{fig3}.  All networks are compared with their random counterparts
(Rand) \cite{maslov2002}.}
\label{fig4}
\end{figure}

To quantify the degree landscapes independently of the layout
dimension, we introduce two measures. First, inspired by the
information horizons in networks
\cite{valverde,trusina2005,rosvallpre2005}, we present a revised
hierarchy-measure ${\cal F}(\ell)$, to estimate the size of the
mountains.  ${\cal F}(\ell)$ is the fraction of pairs of nodes at
distance $\ell$ that are hierarchically connected. Figure
\ref{fig4}(a) shows that ${\cal F}(\ell)$ decreases fastest for the
random-rank hierarchy at a length scale $\ell \approx 4$ corresponding
to the width of the ridge landscape shown in Fig.\ \ref{fig2}(c).
Figure \ref{fig4}(b-d) shows the real-world networks as in Fig.\
\ref{fig3} compared with their random counterparts with the same
degree sequence \cite{maslov2002}. Yeast behaves qualitatively like
the random-rank hierarchy with $\varepsilon$ between $0$ and $0.1$,
which probably reflects some functional localization.  Contrary,
despite their embedding in real space, the Internet and Manhattan both
have a substantial fraction of long degree-hierarchical paths,
corresponding to wide mountains.  However, the randomized counterparts
of the two latter networks, with more peaked mountain landscapes, are
both more degree hierarchical than the real networks.

We define the width of a mountain as the length where 50 procent of
the paths are hierarchical.  Figure \ref{fig4} shows that the average
width of the mountains in the random-rank hierarchy and yeast is about
4. In Manhattan and the Internet it is larger, about 6, and in the
degree-rank hierarchy it is by definition the network diameter.

In the second landscape measure, we measure the separation between
mountains to investigate how the hubs are positioned relative to each
other.  $d(k_{\rm hub})$ is associated to maximum distances between
nodes with degree $k$ equal or larger to the threshold value $k_{\rm
hub}$.  It is defined by the distance from one hub to its most distant
hub in the network, averaged over all hubs
\begin{equation}
d(k_{\rm hub}) = \frac{1}{N_{k\geqslant k_{\rm
hub}}}\sum_{\{i|k_i\geqslant k_{\rm hub}\}}\max_{\{j|k_j\geqslant
k_{\rm hub}\}}d_{ij},
\label{eq.1}
\end{equation}
with $d_{ij}$ being the length of the shortest path between $i$ and
$j$, and $k_i$ the degree of node $i$.  The value of $d(k_{\rm hub})$
is highly dependent of the definition of a hub, and we therefore
measure $d(k_{\rm hub})$ for all values of $k_{\rm hub}$.

\begin{figure}[tp]
\includegraphics[width=\columnwidth]{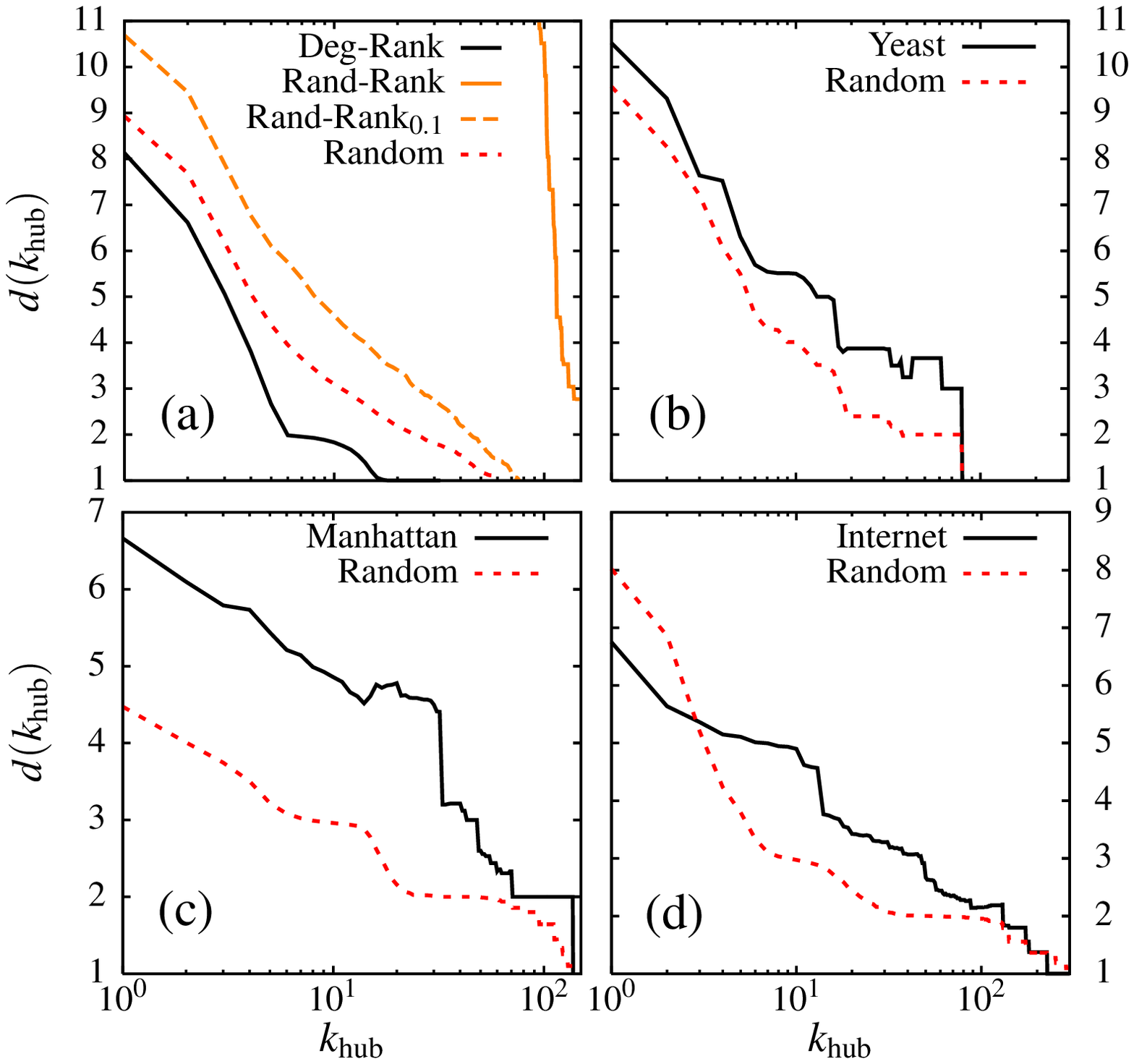}
\caption{ Average longest distance $d(k_{\rm hub})$ between nodes of
degree $k \geqslant k_{\rm hub}$ as a function of $k_{\rm hub}$
(Eq. \ref{eq.1}), for the same networks as in Fig.\ \ref{fig4}.  }
\label{fig5}
\end{figure}

Figure \ref{fig5} shows $d(k_{\rm hub})$ for a few different networks
and their random counterparts. Figure \ref{fig5}(a) shows that the
one-mountain landscapes, the degree-rank hierarchy and the random
network, both have hubs tightly connected.  Contrary, the hubs in the
random-rank hierarchy are extremely separated ($d(1) \approx 100$) all
the way out to a very high hub-threshold value. All the real-world
networks in Fig.\ \ref{fig5}(b-d) fall in between these extremes, but
with a higher $d(k_{\mathrm{hub}})$ than randomly expected for most
values of $k_{\mathrm{hub}}$.  Manhattan, (Fig.\ \ref{fig5}(c)), and
the Internet, (Fig.\ \ref{fig5}(d)), are close to random for really
high degrees, while yeast, (Fig.\ \ref{fig5}(b)), has a separation for
all sizes.  The close resemblance between the random-rank hierarchy
and yeast in Fig.\ \ref{fig4} and \ref{fig5} suggests that the
separation of hubs probably reflects a separation of function at all
scales.

Manhattan is mainly a planned city where the largest hubs,
corresponding to streets and avenues, are connected to each other in a
bipartite way. This results in a $d(k_{\rm hub})$ close to 2 for the
largest hubs.  The Internet is constructed with a hierarchical
structure within each country, and all intermediate-degree nodes
(typically connected to low degree nodes \cite{maslov2002b}) are
therefore separated from each other globally. However, the largest
hubs interconnect the countries, and are therefore connected with each
other.  This results in a $d(k_{\rm hub})$ close to 1 for the largest
hubs.

To summarize, we have generalized the degree-organizational view of
real-world networks with broad degree-distributions, in a landscape
analogue with mountains (high degree nodes) and valleys (low degree
nodes).  To quantify the topology and to be able to compare networks,
we have measured the widths of the mountains and the separation
between different mountains.  We found that the dual map of Manhattan
consists approximately only of one mountain. This implies that typical
navigation between a source and a target in the city involves first
going to larger and larger streets, and then to smaller and smaller
streets. The Internet shares this one-mountain landscape, but the
spatial constraints are weaker and the width of the mountain is about
the same despite the substantially larger network.

Finally, the topological landscape in the protein-interaction network
in yeast has a topology with numerous separated hills.  We suggest
that this reflect functional localization, where proteins tend to be
connected because of similar functions rather than because they have
similar degree.

\end{document}